\documentclass[preprint,aps]{revtex4}

\usepackage{graphicx}

\newcommand{\bq}{\begin{equation}}
\newcommand{\eq}{\end{equation}}
\newcommand{\bqn}{\begin{eqnarray}}
\newcommand{\eqn}{\end{eqnarray}}

\newcommand{\lb}{\label}

\newcommand{\rr}{\bf r}

\newcommand{\myfigure}[3]{
        \begin{figure}
        \centerline{
        \includegraphics{#1.ps}}
        \caption{#2}
        \label{#3}
        \end{figure}
}

\begin{document}

\title{Star Models with Dark Energy}

\author{R. Chan}
\email{chan@on.br}
\affiliation{Coordena\c c\~ao de Astronomia e Astrof\'{\i}sica,
Observat\'orio Nacional, Rua General Jos\'e Cristino 77, S\~ao
Crist\'ov\~ao, CEP 20921-400, Rio de Janeiro, RJ, Brazil}

\author{M. F. A. da Silva}
\email{mfasnic@gmail.com}
\affiliation{Departamento de F\' {\i}sica Te\' orica, Universidade do
Estado do Rio de Janeiro,
Rua S\~ ao Francisco Xavier $524$, Maracan\~ a, 
CEP 20550-013, Rio de Janeiro, RJ, Brazil}

\author{Jaime F. Villas da Rocha}
\email{jaime@mast.br}
\affiliation{Coordenadoria de Educa\c c\~ao em Ci\^encias, Museu de
Astronomia e Ci\^encias Afins,
Rua General Bruce 586, S\~ao Crist\'ov\~ao,
CEP 20921-030, Rio de Janeiro, RJ, Brazil}

\date{\today}

\begin{abstract}

We have constructed star models consisting of four parts: 
(i) a homogeneous inner 
core with anisotropic pressure (ii) an infinitesimal thin shell separating the
core and the envelope; (iii) an envelope of inhomogeneous density and 
isotropic pressure; (iv) an infinitesimal thin shell matching the envelope
boundary and the exterior Schwarzschild spacetime.  We have analyzed all the
energy conditions for the core, envelope and the two thin shells. 
We have found that, in order to have static solutions, at least one of the
regions must be constituted by dark energy.
The results show that there is no physical reason to have a superior
limit for the mass of these objects but for the ratio of mass and radius.

\end{abstract}

\maketitle

\section{Introduction}
Over the past decade, one of the most remarkable discoveries is
that our universe is currently accelerating. This was first
observed from high red shift supernova Ia \cite{agr98}-\cite{Obs5}, and
confirmed later by cross checks from the cosmic microwave
background radiation \cite{wmap}-\cite{wmap1} and large scale structure
\cite{sdss}-\cite{sdss5}.

In Einstein's general relativity, in order to have such an
acceleration, one needs to introduce a component to the matter
distribution of the universe with a large negative pressure. This
component is usually referred as  dark energy. Astronomical
observations indicate that our universe is flat and currently
consists of approximately $2/3$ dark energy and $1/3$ dark
matter. The nature of dark energy as well as dark matter is
unknown, and many radically different models have been proposed,
such as, a tiny positive cosmological constant, quintessence \cite{quint1}-\cite{quint3},
DGP branes \cite{DGP1}-\cite{DGP2}, the non-linear F(R) models \cite{FR1}-\cite{FR3}, 
and dark energy in brane worlds, among many others \cite{other1}-\cite{Brandt07}; see also the review articles \cite{DE1}-\cite{DE2}, and references therein.

On the other hand, another very important issue in gravitational
physics is black holes and their formation in our universe. Although
it is generally believed that on scales much smaller than the horizon
size the fluctuations of dark energy itself are unimportant \cite{Ma99},
their effects on the evolution of matter overdensities may be
significant \cite{FJ97}-\cite{FJ97a}. Then, a natural question is how dark
energy affects the process of the gravitational collapse of a
star. It is known that dark energy exerts a repulsive force
on its surrounding, and this repulsive force may
prevent the star from collapse. Indeed, there are speculations that
a massive star doesn't simply collapse to form a black hole,
instead,  to the formation of stars that contain dark energy.
In a recent work, Mazur and Mottola \cite{Mazur02} have suggested a solution
with a final configuration without neither singularities nor horizons,
which they called "gravastar" (gravitational vacuum star).  In this case,
the gravastar is a system characterized by having a thin shell but not
infinitesimal made of stiff matter, which separates an inner region with
de Sitter spacetime from the Schwarzschild exterior spacetime.  
The elimination of the apparent horizon is done using suitable choice
of the inner and outer radius of the thin shell, in such way that the inner
radius be shorter than the horizon radius of de Sitter spacetime and the
outer radius longer than the Schwarzschild horizon radius.
In a later
work, Visser and Wiltshire \cite{Visser04} have shown that the gravastar is
dynamically stable.  The possibility of the existence of objects like the
gravastar brings all the discussions about the fact that is unavoidable that 
gravitational collapse always forms a black hole.
As a result, black holes may not exist at all \cite{DEStar}-\cite{DEStar1}.
Another related issue is that how dark energy affects
already-formed black hole is related to the fact that 
 it was shown that the mass of a black hole decreases due
to phantom energy accretion and tends to zero when the Big Rip
approaches \cite{BDE04}-\cite{BDE04a}.
Gravitational collapse and formation of black holes in the presence of dark
energy were first considered by several works \cite{BHDE1}-\cite{BHDE4}.

Based on the discussions about the gravastar picture some authors have proposed
alternative models \cite{Horvat}-\cite{Rocha2}. Among them, we can find a Chaplygin dark star \cite{Paramos},
a gravastar supported by non-linear electrodynamics \cite{Lobo07},
a gravastar with continuous anisotropic pressure \cite{CattoenVisser05}. Beside these
ones,
Lobo \cite{Lobo} has studied two models for a dark energy fluid. One of them
describes a homogeneous energy density and another one which uses an
ad-hoc monotonic decreasing energy density, both of them with anisotropic
pressure.  In order to match an exterior Schwarzschild spacetime he has
introduced a thin shell between the interior and the exterior spacetimes.

Since all the works cited previously are based in particular solutions,
the investigation of more general solutions or others particular ones
is important in order to establish the generality of these results.
Our aim with this work is to construct another alternative model to 
black holes considering the possibility of gravitational trapping of dark
energy by standard energy, i.e., not dark energy.  In this context we could
have the dark energy sustaining the collapse of the standard energy, while the
standard matter would trap the dark energy.
In our model the mass function is a
natural consequence of the Einstein's field equations and the energy
density as well as the pressure decreases with the radial coordinate (envelope),
as expected for known stellar models.  In order to eliminate the central
singularity present in this model, we have considered a core with a homogeneous
energy density, described by the Lobo's first solution \cite{Lobo}.
The junction between the envelope and the Schwarzschild exterior spacetime
has imposed a presence of a thin shell, as well as, the junction between
the core and the envelope.

The paper is organized as follows. In Section
II we show the Einstein field equations. In Section III we present
a particular solution that represents an isotropic and inhomogeneous 
dark energy fluid. In order
to construct a more realistic model 
we consider it as  the envelope of the star with a core  constituted by a 
regular anisotropic and homogeneous fluid and a Schwarzschild exterior.
In Section IV we show the junction conditions between the core and envelope
regions and between the envelope and the exterior spacetime. Finally, in 
Section  V we present our final considerations.

\section{The Field Equations}

We consider here a static spherically symmetric spacetime
given by the following metric
\bq
ds^2=-{\rm exp}\left[2\int
g(\tilde{r})d\tilde{r}\right]\,dt^2+\frac{dr^2}{1- 2m(r)/r}
+r^2 (d\theta ^2+\sin ^2{\theta} d\phi ^2), 
\lb{metric}
\eq
where $g(r)$ and $m(r)$ are arbitrary functions of the radial
coordinate, $r$. 

The stress-energy tensor for an isotropic distribution of matter
is given by
\bq
T_{\mu\nu}=(\rho+p)v_\mu v_\nu+p g_{\mu\nu},
\eq
where $v^\mu$ is the four-velocity, $\rho$ is the energy
density, $p$ is the radial pressure measured in the radial direction.

Thus, the Einstein field equation, $G_{\mu\nu}=8\pi T_{\mu\nu}$,
where $G_{\mu\nu}$ is the Einstein tensor, provides the following
relationships
\bqn
m'&=&4\pi r^2 \rho \lb{m},\\
g&=&\frac{m+4\pi r^3 p}{r(r-2m)} \lb{g},\\
p'&=&-\frac{(\rho+p)(m+4\pi r^3 p)}{r(r-2m)} 
\lb{TOV},
\eqn
where the prime denotes a derivative with respect to the radial
coordinate, $r$. Equation (\ref{TOV}) corresponds to the
Tolman-Oppenheimer-Volkoff (TOV) equation.

The standard cosmology considers the Universe constituted of a perfect
fluid described by an equation of state of the type $p=\omega \rho$.
Using this type of equation of state with a sufficiently negative pressure
could explain the positive acceleration of the Universe in the context of the
General Relativity theory.  We consider this kind of equation of state
in order to investigate the conditions of existence of dark energy star.
Assuming an equation of state of the type $p=\omega \rho$ and substituting it
in the equation (\ref{TOV}) we have
\bq
\omega \rho'+(\rho+\omega \rho) \frac{(m+4 \pi r^3 \omega \rho)}{r (r-2 m)}=0.
\lb{TOVw}
\eq

Solving equation (\ref{TOVw}) in terms of $m$ we get
\bq
m = \frac{\omega r^2 \left(\rho'+4 \pi r \omega \rho^2+4 \pi r \rho^2\right)}
{2 \omega \rho' r-\omega \rho-\rho}.
\lb{m2}
\eq

Differentiating equation (\ref{m2}) and substituting into equation (\ref{m})
we obtain
\bqn
& &\frac{r}{(2 \omega \rho' r-\omega \rho-\rho)^2} \left(-24 \pi r^2 \omega^2 \rho^2 \rho'-12 \pi r^2 \rho^2 \omega \rho'+28 \pi r \omega^2 \rho^3+20 \pi r \omega \rho^3+4 \pi r \rho^3 \right. \nonumber \\
&-&\left. 3 \omega^2 \rho'^2 r+2 \rho' \omega^2 \rho+2 \rho' \omega \rho-\omega \rho'^2 r+12 \pi r \omega^3 \rho^3+r \rho'' \omega^2 \rho+8 \pi r^3 \omega^3 \rho^2 \rho'' +8 \pi r^3 \omega^2 \rho^2 \rho'' \right. \nonumber \\
&-& \left. 16 \pi r^3 \omega^3 \rho \rho'^2-12 \pi r^2 \omega^3 \rho^2 \rho'+r \rho'' \omega \rho \right) = 0.
\lb{eq1}
\eqn

Below we work with a particular solution of this equation.

\section{Solution of the Physical Equations}

Solving equation (\ref{eq1}) in terms of $\rho$ we can obtain a particular
solution written as
\bq
\rho = \frac{\omega}{2 \pi (\omega^2+6 \omega+1) r^2},
\lb{rho}
\eq
where we can note that the values of $\omega$ can not be $-3+2\sqrt{2}$ or 
$-3-2\sqrt{2}$, in order to avoid the singularities for these values.  
Since $\rho \ge 0$ thus $-3-2\sqrt{2} < \omega < -3+2\sqrt{2}$ or $\omega\ge0$.
Besides, our envelope with $\omega=1$ reproduces the same rigid
fluid used by Mazur and Mottola \cite{Mazur02}, with the pressure proportional to $r^{-2}$.

Thus, using equations (\ref{m}) and (\ref{g}) we obtain that
\bq
m(r) = \frac{2\omega r}{\omega^2+6\omega+1},
\lb{mr}
\eq
\bq
g(r) = \frac{2\omega}{(\omega+1)r}.
\lb{gr}
\eq

The metric can be rewritten as
\bq
\lb{envelope_metric}
ds_{+}^2=-r^{4\omega/(\omega+1)} dt^2 + \frac{\omega^2+6\omega+1}{(\omega+1)^2} dr^2 + r^2 \left( d\theta^2 + \sin^2 \theta d\phi^2 \right),
\lb{ds2+}
\eq
where the subscript (+) denotes the envelope spacetime as we consider it (see below).
Since the metric function $g_{rr}$ is negative for 
$-3-2\sqrt{2} < \omega < -3+2\sqrt{2}$, this interval is not allowed.
Thus, this solution implies $\omega \ge 0$.

The Kretschmann scalar for (\ref{envelope_metric}) is given by
\bq
K = R_{\alpha \beta \mu \nu} R^{\alpha \beta \mu \nu} = 
\frac{16\omega^2(3\omega^2+2\omega+7)}{r^4(\omega^2+6\omega+1)}
\lb{K},
\eq
where $R_{\alpha \beta \mu \nu}$ is the Riemann tensor.  We can note that 
$\lim_{r \rightarrow \infty} K = 0$, i.e., this
solution is asymptotically flat.  But, we can see also that
$\lim_{r \rightarrow 0} K \rightarrow +\infty$, the solution is singular
at the origin.  Then, we consider the metric (\ref{ds2+}) as an envelope 
solution.  Below we present a core solution.

\subsection{Core Solution}

In order to avoid the singularity at $r=0$ we cut the spacetime (\ref{envelope_metric})  around  its
origin and fit another one, an anisotropic fluid with the density of 
energy $\mu$ constant.  We have chosen an anisotropic solution for the core
since it was shown that gravastars solutions must exhibit anisotropic
pressures to be finite-sized objects \cite{CattoenVisser05}.  Using the results from Ref. \cite{Lobo} 
we have 
\bq
\mu=\mu_0=constant,
\eq
\bq
p_r=k\mu_0, 
\eq
and 
\bq
\lb{ptpr}
p_t=p_r\left[1+\frac{(1+k)(1+3k)M({\rr})}{2k({\rr}-2M({\rr}))}\right].
\eq
In this case, we have as isotropic pressure limits, $k=-1/3$ and $k=-1$,
otherwise we have anisotropic pressures.  
Here we point out that the anisotropy in the pressure changes the ranges 
where the energy conditions are satisfied, which depend on the values of
$M(\rr)/\rr$ \cite{Chan08a}.

The core metric can be written as
\bq
ds_{-}^2=-\left[1-\frac{2M({\rr})}{\rr}\right]^{-(1+3k)/2}dv^2+
\left[ { 1-\frac{2M({\rr})}{{\rr}} } \right]^{-1}d{\rr}^2
+{\rr}^2 (d\theta ^2+\sin ^2{\theta} d\phi ^2),
\lb{ds2-}
\eq
where $M({\rr}) = 4\pi \mu_0 {\rr}^3/3$ and the subscript (-) means
the core spacetime. See figure \ref{core}.

Note from the equation (\ref{ptpr}) that the anisotropy of the core
$|p_t-p_r|$ decreases with the {\rr}
coordinate. Then, the supposition of an isotropic envelope is physically
reasonable.

\myfigure{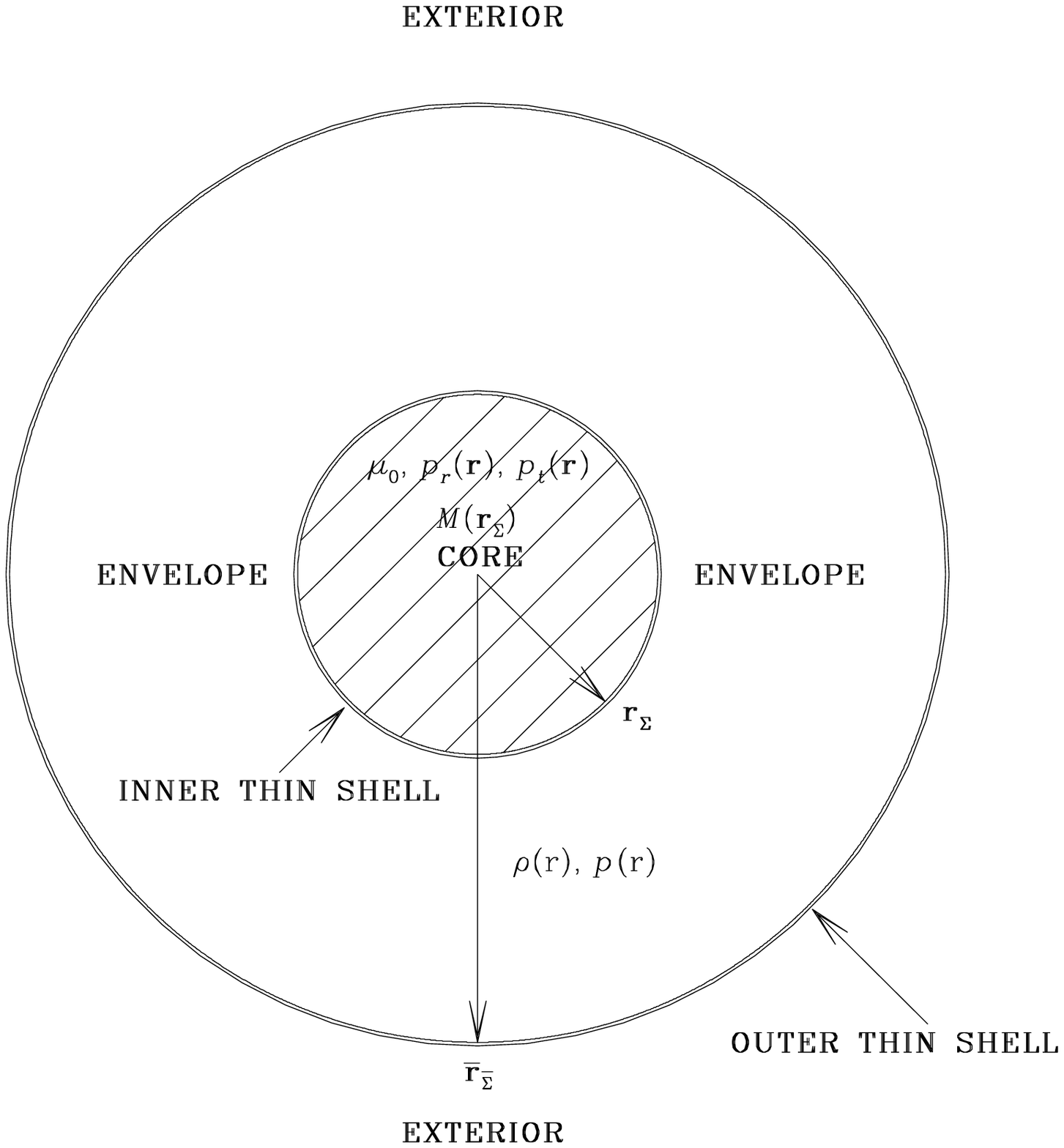}{The regions of the star: the core, the envelope and the
exterior.
The radius of the core is ${\rr}_{\Sigma}$ and the total mass inside
${\rr}_{\Sigma}$ is $M({\rr}_{\Sigma})$. The density  of the core is 
homogeneous and the pressure can be isotropic or anisotropic. The outer radius
of the envelope is $\bar {\rr}_{\bar \Sigma}$.}
{core}

\subsection{Exterior Solution}

In order to limit the matter of the star we match the exterior
spacetime with a Schwarzschild solution.
The Schwarzschild exterior metric can be written as
\bq
ds_{e}^2=-\left(1-\frac{2\bar M}{\bar {\rr}}\right)du^2+
\left( { 1-\frac{2\bar M}{\bar {\rr}} } \right)^{-1}d\bar {\rr}^2
+\bar {\rr}^2 (d\theta ^2+\sin ^2{\theta} d\phi ^2),
\lb{ds2e}
\eq
where $\bar M$ is the total mass of the star, including the mass of the core,
and the subscript (e) denotes the exterior spacetime.

\section{Junction Conditions}

The metric of the hypersurface $\Sigma$ at the frontier of the core and the
envelope is given by
\bq
ds_{\Sigma}^2=-d \tau^2 + R(\tau)^2 (d\theta ^2+\sin ^2{\theta} d\phi ^2),
\eq
where $\tau$ is the time coordinate define only on $\Sigma$.

The metric of the hypersurface $\bar \Sigma$ at the frontier of the envelope 
and the exterior is given by
\bq
ds_{\bar \Sigma}^2=-d \bar \tau^2 + \bar R(\bar \tau)^2 (d\theta ^2+\sin ^2{\theta} d\phi ^2),
\eq
where $\bar \tau$ is the time coordinate define only on $\bar \Sigma$.

From the conditions $(ds_{-}^2)_{\Sigma}=(ds_{+}^2)_{\Sigma}$ and 
$(ds_{+}^2)_{\bar \Sigma}=(ds_{e}^2)_{\bar \Sigma}$ and using
equations (\ref{ds2+}), (\ref{ds2-}) and (\ref{ds2e}), 
we get the following relations
\bq
{\rr}_{\Sigma}=r_{\Sigma}=R(\tau),
\eq
\bq
{\bar {\rr}}_{\bar \Sigma}=r_{\bar \Sigma}=\bar R({\bar \tau}),
\eq
\bq
\left(\frac{dv}{d\tau}\right)^2= \left[ 1 - \frac{2M({\rr}_{\Sigma})}{{\rr}_{\Sigma}} \right]^
{(1+3k)/2},
\eq
\bq
\left(\frac{dt}{d\tau}\right)^2= r_{\Sigma}^{-4\omega/(\omega+1)},
\eq
\bq
\left(\frac{du}{d\bar \tau}\right)^2= \left( 1 - \frac{2\bar M}{\bar {\rr}_{\bar \Sigma}} \right)^{-1},
\eq
\bq
\left(\frac{dt}{d\bar \tau}\right)^2= r_{\bar \Sigma}^{-4\omega/(\omega+1)}.
\eq

The core extrinsic curvature is given by
\bq
K^{-}_{\tau\tau}=-\frac{1}{2{\rr}_{\Sigma}} \left[\frac{-2M({\rr}_{\Sigma})+{\rr}_{\Sigma}}{{\rr}_{\Sigma}}\right]^
{-3k/2} 
\left(\frac{dv}{d\tau}\right)^2 \frac{1}{2 M({\rr}_{\Sigma})-{\rr}_{\Sigma}} 
\left[M({\rr}_{\Sigma})-M'({\rr}_{\Sigma}) {\rr}_{\Sigma}\right](3k+1),
\eq
\bq
K^{-}_{\theta\theta}=-\left[\frac{-2 M({\rr}_{\Sigma})+{\rr}_{\Sigma}}{{\rr}_{\Sigma}}\right]^{-1/2} [2M({\rr}_{\Sigma})-{\rr}_{\Sigma}],
\eq
\bq
K^{-}_{\phi\phi}=-\left[\frac{-2M({\rr}_{\Sigma})+{\rr}_{\Sigma}}{{\rr}_{\Sigma}}\right]^{-1/2} [2 M({\rr}_{\Sigma})-{\rr}_{\Sigma}] 
\sin^2 \theta,
\eq
and the envelope extrinsic curvature is given by
\bq
K^{+}_{\tau\tau}=-2 r_{\Sigma}^{-4/(\omega+1)} r_{\Sigma}^3 \sqrt{\omega^2+6\omega+1} \left(\frac{dt}{d\tau}\right)^2
\frac{1}{\omega^2+6\omega+1} \omega, 
\eq
\bq
K^{+}_{\theta\theta}=\sqrt{\omega^2+6\omega+1} (\omega+1) \frac{1}{\omega^2+6\omega+1} r_{\Sigma},
\eq
\bq
K^{+}_{\phi\phi}=\sqrt{\omega^2+6\omega+1}(\omega+1) \frac{1}{\omega^2+6\omega+1} r_{\Sigma} \sin^2 \theta.
\eq

The core extrinsic curvature can be rewritten as
\bqn
K^{-}_{\tau\tau}&=& \frac{(1+3k)}{2{\rr}_{\Sigma}^2}
\left[ 1 - \frac{2M({\rr}_{\Sigma})}{{\rr}_{\Sigma}} \right]^
{-\frac{1}{2}} \left[ M({\rr}_{\Sigma}) - M'({\rr}_{\Sigma}) {\rr}_{\Sigma} \right] \nonumber \\ &=&
-(1+3k)\frac{M({\rr}_{\Sigma})}{{\rr}_{\Sigma}^2}
\left[ 1 - \frac{2M({\rr}_{\Sigma})}{{\rr}_{\Sigma}} \right]^{-\frac{1}{2}}, 
\lb{Ktautau-}
\eqn
\bq
K^{-}_{\theta\theta}= {\rr}_{\Sigma} \left[ 1 - \frac{2M({\rr}_{\Sigma})}{{\rr}_{\Sigma}} \right]^{\frac{1}{2}}, 
\lb{Kthetatheta-}
\eq
\bq
K^{-}_{\phi\phi}= K^{-}_{\theta\theta} \sin^2 \theta,
\lb{Kphiphi-}
\eq
the envelope extrinsic curvature can be rewritten as
\bq
K^{+}_{\tau\tau}= -\frac{2\omega}{r_{\Sigma}}
\left( \omega^2+6\omega+1 \right)^{-\frac{1}{2}},
\lb{Ktautau+}
\eq
\bq
K^{+}_{\theta\theta}= r_{\Sigma} (\omega+1)
\left( \omega^2+6\omega+1 \right)^{-\frac{1}{2}},
\lb{Kthetatheta+}
\eq
\bq
K^{+}_{\phi\phi}= K^{+}_{\theta\theta} \sin^2 \theta,
\lb{Kphiphi+}
\eq
and the Schwarzschild exterior extrinsic curvature can be written as
\bq
K^{e}_{\bar \tau\bar \tau}= -\frac{\bar M}{\bar {\rr}_{\bar \Sigma}^2}
\left[ 1 - \frac{2\bar M}{\bar {\rr}_{\bar \Sigma}} \right]^{-\frac{1}{2}} 
\lb{Ktautaue}
\eq
\bq
K^{e}_{\theta\theta}= \bar {\rr}_{\bar \Sigma} \left[ 1 - \frac{2\bar M}{\bar {\rr}_{\bar \Sigma}} \right]^{\frac{1}{2}}, 
\lb{Kthetathetae}
\eq
\bq
K^{e}_{\phi\phi}= K^{e}_{\theta\theta} \sin^2 \theta.
\lb{Kphiphie}
\eq

\subsection{Junction conditions between Core-Envelope: without a thin shell}

Using the junction conditions $K^{-}_{\tau\tau}=K^{+}_{\tau\tau}$ and
$K^{-}_{\theta\theta}=K^{+}_{\theta\theta}$ we get two equations
in terms of $M({\rr}_{\Sigma})$ and ${\rr}_{\Sigma}$
\bq
-(1+3k)\frac{M({\rr}_{\Sigma})}{{\rr}_{\Sigma}^2}
\left[ 1 - \frac{2M({\rr}_{\Sigma})}{{\rr}_{\Sigma}} \right]^{-\frac{1}{2}}
+\frac{2\omega}{{\rr}_{\Sigma}}
\left( \omega^2+6\omega+1 \right)^{-\frac{1}{2}}=0,
\eq
and
\bq
{\rr}_{\Sigma} \left[ 1 - \frac{2M({\rr}_{\Sigma})}{{\rr}_{\Sigma}} \right]^{\frac{1}{2}}
- {\rr}_{\Sigma} (\omega+1)
\left( \omega^2+6\omega+1 \right)^{-\frac{1}{2}}=0.
\eq

We can solve these two equations, obtaining the total mass of the core and
the radius of the core as a function of $\omega$ and $k$
giving
\bq
\omega=3k,
\eq
and
\bq
\frac{M({\rr}_{\Sigma})}{{\rr}_{\Sigma}} = \frac{2\omega}{(\omega^2+6\omega+1)} \le \frac{1}{4}.
\lb{MSigmaf}
\eq
Note that although there is a negative interval for
$\omega$ where the mass is positive, this same interval is forbidden
by the metric's signature, equation (\ref{ds2+}). Consequently, $k$ is also 
non negative.  We note that the solution can
represent a star model with an upper limit for the ratio of mass to radius
of the core.

\subsection{Junction conditions between Core-Envelope: with a thin shell}

So, we match a thin shell of energy density $\sigma$ and pressure $P$ at
the frontier between the core and the envelope. Thus, we can write
\bq
\sigma= -\frac{1}{4\pi}\kappa^{\theta}_{\theta},
\eq
\bq
P=\frac{1}{8\pi}\left( \kappa^{\theta}_{\theta}+\kappa^{\tau}_{\tau}
\right),
\eq
where
\bq
\kappa_{ij}= K^+_{ij}-K^-_{ij}.
\eq
Since
\bq
\kappa_{\tau}^{\tau}=
\frac{2\omega}{{\rr}_{\Sigma}}
\left( \omega^2+6\omega+1 \right)^{-\frac{1}{2}}-
(1+3k)\frac{M(\rr_{\Sigma})}{{\rr}_{\Sigma}^2}
\left[ 1 - \frac{2 M(\rr_{\Sigma})}{ {\rr}_{\Sigma}} \right]^{-\frac{1}{2}},
\lb{Ktautau1}
\eq
\bq
\kappa_{\theta}^{\theta}=
\frac{1}{{{\rr}}_{\Sigma}} (\omega+1)
\left( \omega^2+6\omega+1 \right)^{-\frac{1}{2}}-
\frac{1}{{{\rr}}_{\Sigma}} \left[ 1 - \frac{2 M(\rr_{\Sigma})}{{\rr}_{\Sigma}} 
\right]^{\frac{1}{2}},
\lb{Kthetatheta1}
\eq
then
\bq
\sigma=
-\frac{1}{4\pi {\rr}_{\Sigma}}\left\{ 
(\omega+1) \left( \omega^2+6\omega+1 \right)^{-\frac{1}{2}}-
\left[ 1 - \frac{2 M(\rr_{\Sigma})}{{\rr}_ {\Sigma}} \right]^{\frac{1}{2}}
\right\}
\lb{sigma0}
\eq
\bq
P=
\frac{1}{8\pi{\rr}_{\Sigma}} \left\{ 
(3\omega+1) \left( \omega^2+6\omega+1 \right)^{-\frac{1}{2}}-
\left[ 1 + (1+3k)\frac{M(\rr_{\Sigma})}{{\rr}_{\Sigma}} \right]
\left[ 1 - \frac{2M(\rr_{\Sigma})}{{\rr}_{\Sigma}} \right]^{-\frac{1}{2}}
\right\}
\lb{P0}
\eq
where $\omega < -3-2\sqrt{2}$ or $\omega>-3+2\sqrt{2}$ 
(in order to have $\sigma$ and $P$ real) 
and ${\rr}_{\Sigma} > 2 M(\rr_{\Sigma})$. Since $\sigma \ge 0$ and $\rho \ge 0$ 
then $\omega \ge 0$, which represents a not dark energy envelope, but no restriction
is imposed to the core, i.e., we can have any values for $k$. 

\subsubsection{Energy conditions for the thin shell: Core-Envelope}

The energy conditions \cite{HE73} for a thin shell can be written as
\begin{enumerate}
\item Weak: $\sigma \ge 0$ and $\sigma+P \ge 0$
\item Dominant: $\sigma+P \ge 0$ and $\sigma-P \ge 0$
\item Strong: $\sigma+P \ge 0$ and $\sigma+2P \ge 0$
\end{enumerate}

The characterization of dark
energy fluid is the violation of one of the strong energy conditions,
more specifically, that one related the Raychaudhuri equation \cite{Chan08a}.
If the second of the weak energy conditions
is violated, we have a phantom dark energy fluid.

For the inner thin shell, that is, between the core and the envelope, we have 
\bq
\sigma \ge 0
\eq
\bq
\sigma+P = 
\frac{1}{8\pi{\rr}_{\Sigma}} \left\{ 
(\omega-1) \left( \omega^2+6\omega+1 \right)^{-\frac{1}{2}}+
\left[ 1 - (5+3k)\frac{M(\rr_{\Sigma})}{{\rr}_{\Sigma}} \right]
\left[ 1 - \frac{2M(\rr_{\Sigma})}{{\rr}_{\Sigma}} \right]^{-\frac{1}{2}}
\right\}
\eq
\bq
\sigma-P = 
\frac{1}{8\pi{\rr}_{\Sigma}} \left\{ 
-(5\omega+3) \left( \omega^2+6\omega+1 \right)^{-\frac{1}{2}}+
3\left[ 1 - (1-k)\frac{M(\rr_{\Sigma})}{{\rr}_{\Sigma}} \right]
\left[ 1 - \frac{2M(\rr_{\Sigma})}{{\rr}_{\Sigma}} \right]^{-\frac{1}{2}}
\right\}
\eq
\bq
\sigma+2P = 
\frac{1}{4\pi{\rr}_{\Sigma}} \left\{ 
2\omega \left( \omega^2+6\omega+1 \right)^{-\frac{1}{2}}-
3\left[ (1+k)\frac{M(\rr_{\Sigma})}{{\rr}_{\Sigma}} \right]
\left[ 1 - \frac{2M(\rr_{\Sigma})}{{\rr}_{\Sigma}} \right]^{-\frac{1}{2}}
\right\} .
\eq
In the Table I, we summarize the results of the
energy conditions of the inner thin shell for different limits and $k \ge -1/3$.
We analyzed only the case where $k$  $\geq$ -1/3 because we are interested to check 
if it is possible to have all the structures constituted by standard energy.

\begin{table}
\caption{\label{tab:table1}This table summarizes the results of the
energy conditions of the inner thin shell for different limits and $k \ge -1/3$.}
\begin{ruledtabular}
\begin{tabular}{cccc}
Case& $8\pi{\rr}_{\Sigma}(\sigma+P)$ & $8\pi{\rr}_{\Sigma}(\sigma-P)$ & $4\pi{\rr}_{\Sigma}(\sigma+2P)$ \\
\hline
$\omega \rightarrow 0$ & 
$-(4+3k)\frac{M({\rr}_{\Sigma})}{{\rr}_{\Sigma}} \le 0 $ &
$3k\frac{M({\rr}_{\Sigma})}{{\rr}_{\Sigma}} \ge 0 $ &
$-3(1+k)\frac{M({\rr}_{\Sigma})}{{\rr}_{\Sigma}} \le 0 $ \\
$M({\rr}_{\Sigma})/{\rr}_{\Sigma} \ll 1$ & & & \\
\hline
$\omega \rightarrow 0$ &
$-1+\frac{ 1-(5+3k)/2 }{ \sqrt{1-2\frac{ M({\rr}_{\Sigma}) }{ {\rr}_{\Sigma} } } }\le 0 $ & 
$-3+\frac{ 3[1-(1-k)/2] }{ \sqrt{ 1-2\frac{ M({\rr}_{\Sigma}) }{ {\rr}_{\Sigma} } } } \ge 0 $ &
$ \frac{ -3(1+k)/2 }{ \sqrt{ 1- 2\frac{ M({\rr}_{\Sigma}) }{ {\rr}_{\Sigma} } } } \le 0 $  \\
$M({\rr}_{\Sigma})/{\rr}_{\Sigma} \rightarrow 1/2$ & & & \\ 
\hline
$\omega \rightarrow 1$ & 
$1 -(4+3k)\frac{M({\rr}_{\Sigma})}{{\rr}_{\Sigma}} \ge 0 $ &
$-\frac{4}{\sqrt{2}}+ 3\left[1+k\frac{M({\rr}_{\Sigma})}{{\rr}_{\Sigma}} \right] \ge 0 $ &
$\frac{1}{\sqrt{2}} -3(1+k)\frac{M({\rr}_{\Sigma})}{{\rr}_{\Sigma}} \ge 0 $ \\
$M({\rr}_{\Sigma})/{\rr}_{\Sigma} \ll 1$ & 
$\frac{ M({\rr}_{\Sigma}) }{ {\rr}_{\Sigma} } \le \frac{1}{4+3k}$ &
$\frac{M({\rr}_{\Sigma})}{{\rr}_{\Sigma}} \le     \frac{1}{3k}\left(\frac{4}{\sqrt{2}}-3 \right)$ & 
$\frac{M({\rr}_{\Sigma})}{{\rr}_{\Sigma}} \le \frac{1}{3\sqrt{2}(1+k)}$ \\
\hline
$\omega \rightarrow 1$ & 
$\frac{1-(5+3k)/2}{\sqrt{1- 2\frac{M({\rr}_{\Sigma})}{{\rr}_{\Sigma}}}} \le 0 $ &
$-\frac{4}{\sqrt{2}}+\frac{3[1-(1-k)/2]}{\sqrt{1- 2\frac{M({\rr}_{\Sigma})}{{\rr}_{\Sigma}}}} \ge  0 $ &
$\frac{1}{\sqrt{2}}- \frac{3(1+k)/2]}{\sqrt{1- 2\frac{M({\rr}_{\Sigma})}{{\rr}_{\Sigma}}}} \le 0 $  \\
$M({\rr}_{\Sigma})/{\rr}_{\Sigma} \rightarrow 1/2$ & & & 
\end{tabular}
\end{ruledtabular}
\end{table}

\subsection{Junction conditions between Envelope-Exterior: without thin shell}

Using the junction conditions $K^{+}_{\bar \tau\bar \tau}=K^{e}_{\bar \tau\bar \tau}$ and
$K^{+}_{\theta\theta}=K^{e}_{\theta\theta}$ we get two equations
in terms of $\bar M$ and $\bar {\rr}_{\bar \Sigma}$
\bq
-\frac{\bar M}{\bar {\rr}_{\bar \Sigma}^2}
\left[ 1 - \frac{2\bar M}{\bar {\rr}_{\bar \Sigma}} \right]^{-\frac{1}{2}}
+\frac{2\omega}{\bar {\rr}_{\bar \Sigma}}
\left( \omega^2+6\omega+1 \right)^{-\frac{1}{2}}=0,
\lb{Ktautaubar}
\eq
and
\bq
{\bar {\rr}}_{\bar \Sigma} \left[ 1 - \frac{2\bar M}{\bar {\rr}_{\bar \Sigma}} \right]^{\frac{1}{2}}
- {\bar {\rr}}_{\bar \Sigma} (\omega+1)
\left( \omega^2+6\omega+1 \right)^{-\frac{1}{2}}=0.
\lb{Kthetathetabar}
\eq
Considering these two equations we have the
equation
\bq
(\omega+1)^2=1,
\lb{omegabar}
\eq
which gives us
\bqn
\omega&=&0, \nonumber \\
\omega&=&-2,
\eqn
where only the first one is solution of the original equations 
(\ref{Ktautaubar}) and (\ref{Kthetathetabar}). This solution implies that
\bq
\bar M = M({\rr}_{\Sigma}) = 0,
\eq
which means that all the spacetime is Minkowski.  Then we can conclude that
for this kind of fluid distribution, it is not possible to have 
structures surrounded by a Schwarzschild spacetime without the introduction of
a thin shell.

However, in a Lobo's work \cite{Lobo}  
it was necessary the introduction of a thin shell
on the junction hypersurface.  The authors have suggested that it is possible
to match the interior and exterior spacetime without a thin shell. 
However, eliminating the thin shell, vanishing
its energy density and pressure (equations 25 and 26 in the original work), we
get the same junction conditions considered in this present work, i.e. giving
a Minkowski spacetime.  In order to 
do that we assume $m=M$ and $\omega=0$ or $m'=0$ with also $\dot a=0$.
Thus, the unique interior solutions that admits a matching with Schwarzschild 
spacetime, without the introduction of a thin shell,
is the Minkowski ($m'=0$) and the dust ($\omega=0$) solution.

\subsection{Junction Conditions between Envelope-Exterior: with a thin shell}

Since as seen in the above subsections, we can not match an exterior 
Schwarzschild spacetime directly with the
envelope, we build a thin shell of energy density $\bar \sigma$ and pressure
$\bar P$ in order to do this match.
Thus,
\bq
\bar \sigma= -\frac{1}{4\pi}\bar \kappa^{\theta}_{\theta},
\eq
\bq
\bar P=\frac{1}{8\pi}\left( \bar \kappa^{\theta}_{\theta}+\bar \kappa^{\bar \tau}_{\bar \tau}
\right),
\eq
where
\bq
\bar \kappa_{ij}= K^e_{ij}-K^+_{ij}.
\eq
Since
\bq
\bar \kappa_{\bar \tau}^{\bar \tau}=\frac{\bar M}{\bar {\rr}_{\bar \Sigma}^2}
\left( 1 - \frac{2\bar M}{\bar {\rr}_{\bar \Sigma}} \right)^{-\frac{1}{2}}
-\frac{2\omega}{\bar {\rr}_{\bar \Sigma}}
\left( \omega^2+6\omega+1 \right)^{-\frac{1}{2}},
\lb{Ktautaubar1}
\eq
\bq
\bar \kappa_{\theta}^{\theta}=\frac{1}{{\bar {\rr}}_{\bar \Sigma}} \left( 1 - \frac{2\bar M}{\bar {\rr}_{\bar \Sigma}} \right)^{\frac{1}{2}}
- \frac{1}{{\bar {\rr}}_{\bar \Sigma}} (\omega+1)
\left( \omega^2+6\omega+1 \right)^{-\frac{1}{2}},
\lb{Kthetathetabar1}
\eq
then
\bq
\bar \sigma=-\frac{1}{4\pi\bar {\rr}_{\bar \Sigma}}\left[ \left( 1 - \frac{2\bar M}{\bar {\rr}_{\bar \Sigma}} \right)^{\frac{1}{2}}
- (\omega+1)
\left( \omega^2+6\omega+1 \right)^{-\frac{1}{2}}\right],
\lb{sigma}
\eq
\bq
\bar P=\frac{1}{8\pi\bar {\rr}_{\bar \Sigma}}\left[ 
\left( 1 - \frac{\bar M}{\bar {\rr}_{\bar \Sigma}} \right)
\left( 1 - \frac{2\bar M}{\bar {\rr}_{\bar \Sigma}} \right)^{-\frac{1}{2}}
- (3\omega+1)
\left( \omega^2+6\omega+1 \right)^{-\frac{1}{2}}\right],
\lb{P}
\eq
where $\omega < -3-2\sqrt{2}$ or $\omega>-3+2\sqrt{2}$ 
(in order to have $\sigma$ and $P$ real) 
and $\bar {\rr}_{\bar \Sigma} > 2 \bar M$.  Since $\rho \ge 0 $ then the unique
physical system must have $\omega \ge 0$.

Since $\bar \sigma \ge 0$ then we must have 
\bq
\frac{2\omega}{\omega^2+6\omega+1} \le \frac{\bar M}{\bar {\rr}_{\Sigma}} 
\le \frac{1}{2},
\eq
for $\omega\ge 0$, and where
\bq
\frac{2\omega}{\omega^2+6\omega+1} \le  \frac{1}{4}.
\eq

\subsubsection{Energy conditions for the thin shell: Envelope-Exterior}

The energy conditions \cite{HE73} for the thin shell between the 
envelope and the exterior can be written as
\bq
\bar \sigma \ge 0
\eq
\bq
\bar \sigma+\bar P = 
\frac{1}{8\pi\bar {\rr}_{\bar \Sigma}} \left\{ 
-(5\omega+3) \left( \omega^2+6\omega+1 \right)^{-\frac{1}{2}}+
\left[ 3 - 5\frac{\bar M}{\bar {\rr}_{\bar \Sigma}} \right]
\left[ 1 - \frac{2\bar M}{\bar {\rr}_{\bar \Sigma}} \right]^{-\frac{1}{2}}
\right\}
\eq
\bq
\bar \sigma-\bar P = 
\frac{1}{8\pi\bar {\rr}_{\bar \Sigma}} \left\{ 
(\omega-1) \left( \omega^2+6\omega+1 \right)^{-\frac{1}{2}}+
\left[ 1 - 3\frac{\bar M}{\bar {\rr}_{\bar \Sigma}} \right]
\left[ 1 - \frac{2\bar M}{\bar {\rr}_{\bar \Sigma}} \right]^{-\frac{1}{2}}
\right\}
\eq
\bq
\bar \sigma+2\bar P = 
\frac{1}{4\pi\bar {\rr}_{\bar \Sigma}} \left\{ 
-2(2\omega+1) \left( \omega^2+6\omega+1 \right)^{-\frac{1}{2}}+
\left[ 2 - 3\frac{\bar M}{\bar {\rr}_{\bar \Sigma}} \right]
\left[ 1 - \frac{2\bar M}{\bar {\rr}_{\bar \Sigma}} \right]^{-\frac{1}{2}}
\right\} .
\eq

In Table II we 
summarize the results of the
energy conditions of the outer thin shell for different limits.

\begin{table}
\caption{\label{tab:table2}This table summarizes the results of the
energy conditions of the outer thin shell for different limits.}
\begin{ruledtabular}
\begin{tabular}{cccc}
Case& $8\pi{\rr}_{\Sigma}(\bar \sigma+\bar P)$ & $8\pi{\rr}_{\Sigma}(\bar \sigma-\bar P)$ & $4\pi{\rr}_{\Sigma}(\bar \sigma+2\bar P)$ \\
\hline
$\omega \rightarrow 0$ & 
$-2\frac{\bar M}{\bar {\rr}_{\bar \Sigma}} \le 0 $ &
$-2\frac{\bar M}{\bar {\rr}_{\bar \Sigma}} \le 0 $ &
$-\frac{\bar M}{\bar {\rr}_{\bar \Sigma}} \le 0 $ \\
$\bar M/\bar {\rr}_{\bar \Sigma} \ll 1$ & & $ $ & \\
\hline
$\omega \rightarrow 0$ &
$-3+\frac{ 1/2 }{ \sqrt{1-2\frac{ \bar M }{ \bar {\rr}_{\bar \Sigma} } } }\ge 0 $ & 
$-1-\frac{ 1/2 }{ \sqrt{ 1-2\frac{ \bar M }{ \bar {\rr}_{\bar \Sigma} } } } \le 0 $ &
$-2-\frac{ 1/2 }{ \sqrt{ 1- 2\frac{ \bar M }{ \bar {\rr}_{\bar \Sigma} } } } \ge 0 $  \\
$\bar M/\bar {\rr}_{\bar \Sigma} \rightarrow 1/2$ & & & \\ 
\hline
$\omega \rightarrow 1$ & 
$\frac{ \bar M }{ \bar {\rr}_{\bar \Sigma} } \le \frac{3\sqrt{2}-4}{2\sqrt{2}}$ &
$\frac{\bar M}{\bar {\rr}_{\bar \Sigma}} \le     \frac{1}{2}$ &
$\frac{2\sqrt{2}-3}{\sqrt{2}} -\frac{\bar M}{\bar {\rr}_{\bar \Sigma}} \le 0 $ \\
$\bar M/\bar {\rr}_{\bar \Sigma} \ll 1$ & & &  \\
\hline
$\omega \rightarrow 1$ & 
$-\frac{4}{\sqrt{2}}+\frac{1/2}{\sqrt{1- 2\frac{\bar M}{\bar {\rr}_{\bar \Sigma}}}} \ge 0 $ &
$-\frac{1/2}{\sqrt{1- 2\frac{\bar M}{\bar {\rr}_{\bar \Sigma}}}} \le  0 $ &
$-\frac{3}{\sqrt{2}}+ \frac{1/2}{\sqrt{1- 2\frac{\bar M}{\bar {\rr}_{\bar \Sigma}}}} \ge 0 $  \\
$\bar M/\bar {\rr}_{\bar \Sigma} \rightarrow 1/2$ & & & 
\end{tabular}
\end{ruledtabular}
\end{table}

Comparing the Tables I and II we can have the following conclusions:
\begin{enumerate}
\item Limits $\omega \rightarrow 0, \; \frac{M({\rr}_{\Sigma})}{{\rr}_{\Sigma}} \ll 1, \;
\frac{\bar M}{\bar {\rr}_{\Sigma}} \ll 1$: both (inner and outer) thin shells are phantom; 
\item Limits $\omega \rightarrow 0, \; \frac{M({\rr}_{\Sigma})}{{\rr}_{\Sigma}} \rightarrow 1/2, \;
\frac{\bar M}{\bar {\rr}_{\bar \Sigma}} \rightarrow 1/2$: the inner thin shell is made of dark energy and the outer thin shell violates the dominant energy condition;
\item Limits $\omega \rightarrow 1, \; \frac{M({\rr}_{\Sigma})}{{\rr}_{\Sigma}} \ll 1, \; \frac{\bar M}{\bar {\rr}_{\bar \Sigma}} \ll 1$: the inner thin shell is constituted by dark and not dark energy, depending on $\frac{M({\rr}_{\Sigma})}{{\rr}_{\Sigma}}$ and $\frac{\bar M}{\bar {\rr}_{\bar \Sigma}}$, and the outer thin shell is made of dark energy; 
\item Limits $\omega \rightarrow 1, \; \frac{M({\rr}_{\Sigma})}{{\rr}_{\Sigma}} \rightarrow 1/2, \;
\frac{\bar M}{\bar {\rr}_{\bar \Sigma}} \rightarrow 1/2$: the inner thin shell is made of dark energy and the outer thin shell is constituted by not dark energy, violating the dominant energy condition. 
\end{enumerate}

We can conclude with these limits that we always have one or both thin shells
constituted by dark energy.

\section{Conclusions}

We have constructed a star model consisting of four parts: 
(i) a homogeneous inner 
core with anisotropic pressure (ii) an infinitesimal thin shell separating the
core and the envelope; (iii) an envelope of inhomogeneous density and 
isotropic pressure; (iv) an infinitesimal thin shell matching the envelope
boundary and the exterior Schwarzschild spacetime.  We have analyzed all the
energy conditions for the core, envelope and the two thin shells. 

In our model the mass function is a
natural consequence of the Einstein's field equations and the energy
density as well as the pressure decreases with the radial coordinate (envelope),
as expected for known stellar models.  In order to eliminate the central
singularity present in this model, we have considered a core with a homogeneous
energy density, described by the Lobo's first solution \cite{Lobo}.

We have proposed in this work an alternative model and a generalization
to gravastars.  Note that for $k=-1$ we get a vacuum core with a cosmological
constant $\Lambda=8\pi \mu_0$, i.e., a de Sitter solution.  Thus, we have
constructed models with the same structures of the Mazur and Mottola's one \cite{Mazur02}, with
five regions (an interior de Sitter solution + an infinitesimal shell +
a non infinitesimal shell + infinitesimal shell + an exterior Schwarzschild 
solution).  However, in our models each one of these regions is made of
more general kind of fluids.

In the Table III we summarize the results of the
match of the core's spacetime with envelope's spacetime
and the results of the
match of the envelope's spacetime with the exterior spacetime.

Combining the results of the Tables I, II and III we can see that
in the analyzed cases we always have the presence of the dark energy
at least in one of the thin shell or in the core.  Note also that
from the results that there is no physical reason to have a superior
limit for the mass of these objects but for the ratio of mass and radius,
in order to find out which one is made of dark energy.

\begin{table}
\caption{\label{tab:table3}This table summarizes the results of the
match of the core's spacetime with envelope's spacetime
and the results of the
match of the envelope's spacetime with the exterior spacetime.
DEC means Dominant Energy Condition.
We have standard energy core and envelope, but in all the cases we have
at least one of thin shells made of dark energy.}
\begin{ruledtabular}
\begin{tabular}{|c|c|c|c|c|}
Core& Thin Shell:   & Envelope  & Thin Shell:       & Exterior \\
    & Core-Envelope &           & Envelope-Exterior & Spacetime\\

\hline

\hline
Minkowski & None & Minkowski & None & Minkowski \\

\hline
          & $\omega \rightarrow 0$,\,$\frac{\bar M}{\bar {\rr}_{\bar \Sigma}} \ll 1$ & & $\omega \rightarrow 0$,\,$\frac{\bar M}{\bar {\rr}_{\bar \Sigma}} \ll 1$ & \\
 & Phantom &                 & Phantom &              \\

 & & & & \\

          & $\omega \rightarrow 0$,\,$\frac{\bar M}{\bar {\rr}_{\bar \Sigma}} \rightarrow 1/2 $ & & $\omega \rightarrow 0$,\,$\frac{\bar M}{\bar {\rr}_{\bar \Sigma}} \rightarrow 1/2$ & \\
 & Dark Energy &                 & Not Dark Energy &              \\
          & & & Violation of DEC & \\

Not Dark Energy  & & $\omega > 0$& & Schwarzschild\\

          & $\omega \rightarrow 1$,\,$\frac{\bar M}{\bar {\rr}_{\bar \Sigma}} \ll 1$ & Not Dark Energy& $\omega \rightarrow 0$,\,$\frac{\bar M}{\bar {\rr}_{\bar \Sigma}} \ll 1$ & \\
     & Dark or &                 & Dark Energy &              \\
          & Not Dark Energy & & & \\

 & & & & \\

          & $\omega \rightarrow 1$,\,$\frac{\bar M}{\bar {\rr}_{\bar \Sigma}} \rightarrow 1/2 $ & & $\omega \rightarrow 0$,\,$\frac{\bar M}{\bar {\rr}_{\bar \Sigma}} \rightarrow 1/2 $ & \\
                & Dark Energy &                 & Not Dark Energy &              \\
          &  & & Violation of DEC & \\

\end{tabular}
\end{ruledtabular}
\end{table}

\section*{Acknowledgments}

We would like to thank Dr. Anzhong Wang for the helpful suggestions.
The financial assistance from UERJ
(JFVdaR) and FAPERJ/UERJ (MFAdaS) is gratefully acknowledged. The
author (RC) acknowledges the financial support from FAPERJ (no.
E-26/171.754/2000, E-26/171.533/2002 and E-26/170.951/2006).  
The authors (RC and MFAdaS) also acknowledge the financial support from 
Conselho Nacional de Desenvolvimento Cient\'{\i}fico e Tecnol\'ogico - 
CNPq - Brazil.  The author (MFAdaS) acknowledges the financial support
from Financiadora de Estudos e Projetos - FINEP - Brazil (Ref. 2399/03).

\end{document}